\documentstyle[12pt]{article}
\makeatletter
\@addtoreset{equation}{section}
\makeatother
\renewcommand{\theequation}{\thesection.\arabic{equation}}
\def\lb{\hfil\penalty-10000}

\def\ie{{\sl i.e.\/}}

\begin{document}
\thispagestyle{empty}
 \mbox{} \hspace{1.0cm}February 1993 \hspace{6.4cm}HLRZ-93-08\\
 \mbox{} \hspace{3.5cm} \hspace{7.1cm}BI-TP 93/04\\
\begin{center}
\vspace*{1.0cm}
{{\large Vacuum Tunneling and Periodic Structure in  \\
Lattice Higgs Models}
 } \\
\vspace*{1.0cm}
{\large F.~Karsch$^{1,2}$,
        M.~L.~Laursen$^{1,2}$ \\
        T.~Neuhaus$^2$ and B.~Plache$^2$} \\
\vspace*{1.0cm}
{\normalsize
$\mbox{}^1$ {HLRZ, c/o KFA J\"{u}lich,
             P.O. Box 1913, D-5170 J\"{u}lich, Germany}\\
$\mbox{}^2$ {Fak. f. Physik, Univ. Bielefeld,
              D-4800 Bielefeld, Germany}}\\
\vspace*{2cm}
{\large \bf Abstract}
\end{center}
\setlength{\baselineskip}{1.3\baselineskip}

Using a geometric definition  for the lattice Chern-Simons term in even
dimensions, we have studied the distribution of Chern-Simons numbers for the
2d-U(1) and the 4d-SU(2) lattice Higgs models. The periodic structure
of the distributions is preserved in our lattice formulation and has been
examined in detail. In both cases the finite size effects visible in
the distribution of Chern-Simons numbers are well accounted for by the Haar
measure. Moreover, we find that $\langle N_{CS}^2 \rangle$ grows with the
spatial volume. We also find numerical evidence that tunneling in 4d is
increased at high temperature.
\newpage
\setcounter{page}{1}
\section{Introduction}

Recent interest in the finite temperature Electroweak phase transition
has focused on the role of tunneling between topologically distinct
vacuums. Such tunnelings, in a semiclassical approximation,
are well known in the continuum and
lead to a periodic structure in the effective potential \cite{Raja}.
During such a tunneling
the so-called topological charge will change by an integer.
In two dimensions there
are the kinks, which are time independent finite energy
solutions of the scalar Higgs model  or the Sine-Gordon model.
The vacuum solutions occur at spatial infinity and
the tunneling is between $\phi (x = -\infty )$ and $\phi (x = +\infty )$,
$\phi$ being the Higgs field.  In the 2d - Abelian gauge Higgs
system the tunneling is in Euclidean time and governed by time
dependent finite energy solutions -- the vortices.
The tunneling is most transparent if one assumes an axial gauge
$A_{0} = 0$. The vacuum gauge field at temporal infinity becomes pure
gauge $A_{i}(t,x) = \phi (t,x)^{-1}\partial _{i}\phi (t,x)$ and the
tunneling goes from  $A_{i}(t = -\infty ,x)$ to $A_{i}(t = +\infty ,x)$.
In four dimensions one has instantons in the pure SU(2) gauge theory.
This case is very similar to the previous one, upon replacing the
Higgs field with a proper gauge transformation.
With Higgs fields included no time dependent finite energy
solutions are known. However, assuming  static fields only,
one has a saddle point solution -- the
sphaleron \cite{Klink}. The sphaleron in itself does not provide the
tunneling. It is so to speak lying midways between two vacuums and we
must still imagine a time dependent field configuration
interpolating among the two vacuums.

The physical relevance of these tunnelings was pointed out by
't Hooft who found that neither baryon  nor
lepton numbers are conserved in the electroweak theory \cite{Hooft}.
For the baryon and lepton currents one has
\begin{equation}
\partial_{\mu}J_{\mu}^{B} = \partial_{\mu}J_{\mu}^{e} =
 \frac{N_f}{16\pi^{2}} tr[F_{\mu\nu}\tilde{F}_{\mu\nu}],
\end{equation}
where $N_f$ is the number of families of quarks and leptons.
While the $B-L$ symmetry remains unbroken
due to the anomaly cancellation, $B+L$ is no longer conserved.
This so-called baryon number violation is
caused by the nontrivial topological winding of the SU(2) gauge fields.
The baryon number $B$ changes by an amount
\begin{equation}
 \Delta B = \frac{N_f}{16\pi^{2}}
            \int_{t_1}^{t_2} dt \int d^{3}x
            tr[F_{\mu\nu}\tilde{F}_{\mu\nu}].
\end{equation}
In the axial gauge, $A_{0} = 0$, $\Delta B$ is
related to the change in the Chern-Simons(CS) number $N_{CS}$,
\begin{equation}
            \Delta B = N_{f}[N_{CS}(t_2)-N_{CS}(t_1)]
\end{equation}
with
\begin{equation}
           N_{CS}= - \frac{1}{8\pi^{2}} \int d^{3}x
   \epsilon_{ijk} tr[A_{i}(\partial_{j}A_{k}+\frac{2}{3}A_{j}A_{k})].
\end{equation}
At zero temperature only quantum tunneling via instanton like
configurations is possible.
The rate for such a process is, however, exponentially suppressed.
This is because the relevant field
configurations have an action of the order of the barrier
height $2\pi/\alpha_W$, with $\alpha_W$ denoting the electroweak coupling
constant. At high temperatures
such an exponential suppression is absent, since tunneling  can
occur classically  by thermal fluctuations. Assuming that
the temperature is so high that only static fields (kinks or sphalerons)
are relevant, one can go over to a Hamiltonian formulation of the theory and
study the evolution of the system in real time
via the classical Hamiltonian equation of motion or the Langevin equations
with a friction term. Classical nucleation theory can then be applied to
extract information about the rate of baryon number violating processes.
This has been done for the above mentioned lattice models
\cite{Grig,Ambjorn,Shap,Farakos}.
However, a priori it is not known whether the static field approximation
is valid and we, therefore have
attempted to study the full theory in Euclidean time.
By varying the time extent of the lattice
we can in principle control the temperature.
We have initiated work in the 2d - Abelian gauge Higgs model and the
4d - SU(2) gauge Higgs model \cite{Karsch}.

In two dimensions it is easy to derive an expression for the
Chern-Simons number, $N_{CS}$, on the lattice, it amounts to evaluating the
sum of the link angles in a Polyakov line.
In four dimensions this is much more involved. There are several
geometric definitions of $N_{CS}$ \cite{Luesch,Sei,Fox,Gock}.
We decided to use the version by Seiberg \cite{Sei}, which shares
the properties of the continuum  expression.
It is gauge dependent  and changes by an integer under
large gauge transformations. We note that $N_{CS}$ is in general non-integer;
only for configurations which behave as pure gauge it is indeed an integer.
Numerical investigations of topological properties of lattice gauge theories
are known to be difficult and time consuming, if one attempts to preserve
the important quantization properties of topological objects also on the
lattice.  In the geometrical approaches an interpolation of the original
gauge fields into the elementary cells of the d-dimensional lattice is
required and a d-dimensional integration
over the fields needs to be performed. Starting from an existing program for
the calculation of topological charges \cite{Fox} we developed a program for
the calculation of Chern-Simons numbers, which is fully vectorized and very
efficient.

A central goal in studies of CS-numbers in Higgs models on Euclidean
lattices is the calculation of the distribution of CS-numbers,
$P(N_{CS})$, and the determination of
the temperature dependence of the barrier height
between topologically distinct vacuums. In the high temperature limit, where
static field configurations dominate, this is related to the
difference between the minima and maxima of the effective potential,
$V(N_{CS})$, for the CS-numbers, which can be extracted from the
probability distribution,
\begin{eqnarray}
P(N_{CS}) & = & \exp(-V(N_{CS})) \nonumber \\
&  = & \int dAd\phi \exp(-S(A,\phi)) \nonumber \\
                 &    & \delta (N_{CS}+ \frac{1}{8\pi^{2}} \int d^{3}x
\epsilon_{ijk} tr[A_{i}(\partial_{j}A_{k}+\frac{2}{3}A_{j}A_{k})])~~~.
\end{eqnarray}

However, it also is known that even in the continuum at zero temperature,
vacuum fluctuations may lead to rather large values of the CS-numbers and may
easily dominate the above distribution and cover any expected periodic
structure of the potential expected to arise from topological properties of
the field configurations \cite{Ambjorn,Shap,Farakos}.
On the lattice one encounters in addition the problem of finding
a formulation for the CS-numbers which correctly reproduces the topological
properties known in the continuum.
It is the purpose of this paper to systematically
investigate these problems on the lattice. We will study the properties of a
lattice version of the CS-term suggested by Seiberg \cite{Sei}.

The outline of the paper is as follows. In Section 2 we discuss
topological charge and Chern-Simons numbers in the continuum.
In Section 3 we give the
definitions of the lattice topological charge and Chern-Simons term.
In Section 4 we derive expressions for the distribution
of $N_{CS}$ in two dimensions at strong coupling and
we present results for the periodic structure.
Section 5 is devoted to our results in four dimensions,
and we describe the algorithm used in our simulations of the SU(2) Higgs model.
Finally we give our conclusions in Section 6.

\section{Topological charge and the Chern-Simons term in the continuum}
We will first consider the 2d - U(1) case. It is well known that
one can define  a gauge invariant  topological charge $Q$:
\begin{eqnarray}
 Q & = &- \frac{1}{4\pi} \int_{M} dtdx
 \epsilon_{\mu\nu}F_{\mu\nu} \in Z. \nonumber \\
  F_{\mu\nu} & = & \partial_{\mu}A_{\nu} - \partial_{\nu}A_{\mu}
\end{eqnarray}
The manifold is denoted $M$ and we will assume that its boundary
$\partial M$ is a one sphere $S^1$.
The topological charge density, $q$, can be written
as  a divergence of the Chern-Simons density $K_{\mu}$,
\begin{eqnarray}
 q & = & - \frac{1}{4\pi}\epsilon_{\mu\nu}F_{\mu\nu}
 = \partial_{\mu}K_{\mu} \nonumber~~, \\
 K_{\mu} & = & - \frac{1}{2\pi} \epsilon_{\mu\nu} A_{\nu}~~.
\end{eqnarray}
Under a local gauge transformation, $g$, the gauge field, $A_{\mu}$,
changes like
\begin{equation}
 \delta A_{\mu} = \frac{1}{i}g^{-1} \partial_{\mu}g~~,
\end{equation}
  so that
\begin{equation}
 \delta K_{\mu} = - \frac{1}{2\pi i}  \epsilon_{\mu\nu}
                   \partial_{\nu}g\,g^{-1}.
\end{equation}
We define the Chern-Simons number $N_{CS}$ at a given Euclidean time
as follows:
\begin{equation}
 N_{CS} = \int_{\partial M} dx K_{0}.
\end{equation}
We note that
 $N_{CS}$ is an integer only for pure gauge (vacuum) configurations
and is gauge dependent. It changes, however, under a gauge transformation
only by an integer
\begin{equation}
 \delta N_{CS} = - \frac{1}{2\pi i}
  \int_{\partial M} dx  \partial_{1}g\,g^{-1} \in Z.
\end{equation}
This follows also from homotopy theory using the mapping
$g: S^{1} \rightarrow U(1) = S^{1}$.
Such mappings are characterized by the homotopy class
$\Pi_{1}(S^{1}) \in Z$. In the tunneling picture vortices will
interpolate between two vacuums with integer $N_{CS} = n,n+1$.

In the 4d - SU(2) case the topological charge $Q$ is
\begin{eqnarray}
 Q & = & - \frac{1}{32\pi^2} \int_{M} dtd^{3}x
 \epsilon_{\mu\nu\rho\sigma}tr[F_{\mu\nu}F_{\rho\sigma}] \in Z~~,
                                             \nonumber \\
 F_{\mu\nu} & = & \partial_{\mu}A_{\nu} - \partial_{\nu}A_{\mu}
      + [A_{\mu},A_{\nu}]~~.
\end{eqnarray}
Here we shall assume that $\partial M = S^3$.
Like in two dimensions we can write
\begin{eqnarray}
 q & = &- \frac{1}{32\pi^2}
 \epsilon_{\mu\nu\rho\sigma}tr[F_{\mu\nu}F_{\rho\sigma}]
        =  \partial_{\mu}K_{\mu} \nonumber \\
 K_{\mu} & = &- \frac{1}{8\pi^{2}} \epsilon_{\mu\nu\rho\sigma}
   tr[A_{\nu}(\partial_{\rho}A_{\sigma}+
   \frac{2}{3}A_{\rho}A_{\sigma})].
\end{eqnarray}
Under a local gauge transformation, $g$, the gauge field changes as
\begin{equation}
 \delta A_{\mu} =  g^{-1}[A_{\mu} + \partial_{\mu}]g~~,
\end{equation}
giving
\begin{eqnarray}
 \delta K_{\mu} = & - & \frac{1}{24\pi^2}
  \epsilon_{\mu\nu\rho\sigma}
  tr[\partial_{\nu}g\,g^{-1}\, \partial_{\rho}g\,g^{-1}\,
  \partial_{\sigma}g\,g^{-1}] \nonumber \\
                                   & - & \frac{1}{8\pi^2}
  \epsilon_{\mu\nu\rho\sigma}\partial_{\nu}
  tr[\partial_{\nu}g\,g^{-1}\, A_{\sigma}]~~.
\end{eqnarray}
The (timelike) Chern-Simons  number $N_{CS}$ is then defined exactly
as in eq.(2.5), and has the same properties.
Its gauge variation is an integer (the boundary term vanishes)
\begin{equation}
 \delta N_{CS} = - \frac{1}{24\pi^2} \epsilon_{0\nu\rho\sigma}
  \int_{\partial M} d^{3}x
  tr[\partial_{\nu}g\,g^{-1}\, \partial_{\rho}g\,g^{-1}\,
  \partial_{\sigma}g\,g^{-1}] \in Z.
\end{equation}
This time the mapping $g: S^{3} \rightarrow SU(2) = S^{3}$,
and the homotopy class is  $\Pi_{3}(S^{3}) \in Z$. In this case the
instanton will interpolate between two vacuums, assuming the axial gauge.

\section{Topological charge and the Chern-Simons term on the lattice}

We will now consider the lattice version  of the topological charge
and the  Chern-Simons number. We will use a geometric definition given by
Seiberg \cite{Sei}.
Problems with dislocations will be ignored here.
It is convenient to start from the definition of the topological
charge first given by L\"uscher \cite{Luesch}.
The following considerations apply to 2d or 4d.
The manifold $M$ is a torus  and we will  cover it with cells $c(n)$,
where $n$ denotes the lattice sites.
Let the  gauge potential
$A_{\nu}^{n}$ be defined on $c(n)$ and likewise
$A_{\nu}^{n-\hat{\mu}}$ be defined on $c(n-\hat{\mu})$.
At the faces  $f(n,\mu) = c(n-\hat{\mu}) \cap c(n)$, we can relate
the two potentials by a transition function  $v_{n,\mu}(x)$,
\begin{equation}
  A_{\nu}^{n-\hat{\mu}}(x)  =
  v^{-1}_{n,\mu}(x) [A_{\nu}^{n}(x) + \partial_{\nu}] v_{n,\mu}(x).
\end{equation}
At the corners of the faces the transition function is given by
\begin{equation}
 v_{n,\mu}(x) = w^{n-\mu}(x) w^{n}(x)^{-1}~~.
\end{equation}
Here $w^{n}(x)$ is a parallel transporter, used to gauge fix the links
to the complete axial gauge in each cell.
By interpolation this formula is extended to the whole face. The
transition function, $v_{n,\mu} (x)$, defines a bundle, while $w^{n}(x)$ (given
on the boundary of the cell) is a section of the bundle \cite{Goc1}.
In L\"uschers approach \cite{Luesch} the interpolating fields in the interior
of the cells are given by
\begin{eqnarray}
 v_{n,\mu}(x)& = & s^{n-\mu}_{n,\mu}(x)^{-1}v_{n,\mu}(n)s^{n}_{n,\mu}(x),
 \nonumber \\
 s^{t}_{n,\mu}(x) & = & w^{t}(n) S_{n,\mu}(x)w^{t}(x)^{-1},
 \; t=n,n-\mu~~.
\end{eqnarray}
{}From now on we shall assume that $t = n$ or $t = n - \mu$ and that
$\nu \neq \mu$.
In 2d one has explicitly
\begin{eqnarray}
 s^{t}_{n,\mu}(x) & = & [w^{t}(n) U(n,\nu )w^{t}(n+\nu )^{-1}]^{x},
   \nonumber \\
 w^{t}(x) & = & [w^{t}(n) w^{t}(n+\nu )^{-1}]^{x}
 w^{t}(n), \;\;\nu \neq \mu ,\; t=n,n-\mu  ,  \nonumber \\
 S_{n,\mu}(x) & = & U(n,\nu )^{x}.
\end{eqnarray}
The approach of Seiberg is quite similar \cite{Sei}. He
does not apply any axial gauge fixing, and his function
$S_{n,\mu}$ depends only on the original fields. However, otherwise
the interpolation formulas for  $s^{t}_{n,\mu}(x)$ and $S_{n,\mu}(x)$ are
identical. Although the 4d case is more involved the previous statement
holds as well.
For the L\"{u}scher charge \cite{Luesch} one has
\begin{eqnarray}
Q^{L} & = & \sum_{n} q^{L}(n)   =
             \sum_{n,\mu}(-1)^{\mu}(k_{n,\mu} - k_{n+\mu,\mu})
 ,\;\; |q^{L}(n)| \leq 1/2, \nonumber \\
 (-1)^{\mu} k_{t,\mu} & = & -
 \frac{1}{2\pi i} \epsilon_{\mu\nu} \int_{f(t,\mu )} dx
 s^{n}_{t,\mu}(x)^{-1}\partial_{\nu} s^{n}_{t,\mu}(x).
\end{eqnarray}
Notice that $k_{t,\mu}$ is gauge invariant.
After a little algebra one finds
\begin{equation}
 q^{L}(n) = \frac{1}{2\pi i}
 \log [U(n,1)U(n+1,2)U(n+2,1)^{-1}U(n,2)^{-1}]
 ,\;\; |q^{L}(n)| \leq 1/2.
\end{equation}
The  Seiberg charge \cite{Sei} is obtained by replacing
\begin{equation}
 [s^{n}_{t,\mu}(x), k_{t,\mu}]
\rightarrow
 [S_{t,\mu}(x), K_{t,\mu}]~~,
\end{equation}
where $K_{t,\mu}$ can now be interpreted as the local Chern-Simons term.
\begin{eqnarray}
Q^{S} & = & \sum_{n} q^{S}(n)   =
   \sum_{n,\mu}(-1)^{\mu}(K_{n,\mu} - K_{n+\mu,\mu})
 ,\;\; |q^{S}(n)| \leq 1/2,
   \nonumber \\
 (-1)^{\mu} K_{t,\mu} & = &  -
 \frac{1}{2\pi i} \epsilon_{\mu\nu} \int_{f(t,\mu )} dx
 S_{t,\mu}(x)^{-1}\partial_{\nu} S_{t,\mu}(x)~~.
\end{eqnarray}
Though each  $K_{n,\mu}$ term is gauge dependent, the charge remains
gauge invariant, in fact $q^{S}(n) = q^{L}(n)$.
This follows from the relation
\begin{eqnarray}
q^{S}(n) & = & q^{L}(n) - q^{w}(n)~~~, \nonumber \\
  q^{w}(n) & = &
 \frac{1}{2\pi i} \epsilon_{\mu\nu} \int_{\partial c(n)} dx
 w^{n}(x)^{-1}\partial_{\nu} w^{n}(x)~~~.
\end{eqnarray}
The last piece is an integer since we integrate the section over the
boundary of the cell. By restricting the
charge to the interval $|q(n)| \leq 1/2 $, it follows that the two
charge definitions agree.
The Chern-Simons term for the 2d-U(1) Higgs model is now given by
\begin{equation}
 N_{CS} \equiv \sum_{n_s} K_{n_{s},\mu} =
 \frac{1}{2\pi i}\sum_{n_s} \log  U(n_{s},\nu )~~.
\end{equation}
 The summation is only over the spatial lattice at a fixed
Euclidean time, {\it i.e.} $N_{CS} \equiv N_{CS} (t)$. Under gauge
transformations $N_{CS}$ indeed
changes by an integer. Moreover, it is an integer for pure gauge.

We can extend these considerations to the four dimensional case.
Defining for shortness
\begin{eqnarray}
  {\cal S}^{t}_{\nu}(x) & = & s^{n}_{t,\mu}(x)^{-1}
\partial_{\nu}s^{n}_{t,\mu}(x)~~~,
  \nonumber \\
  {\cal P}^{t}_{\nu}(x) & = & p^{n}_{t+\nu,\mu\nu}(x)^{-1}\partial_{\nu}
                    p^{n}_{t+\nu,\mu\nu}(x)~~~,
\end{eqnarray}
one finds
\begin{eqnarray}
Q^{L} & = & \sum_{n} q^{L}(n)   =
             \sum_{n,\mu}(-1)^{\mu}(k_{n,\mu} - k_{n+\mu,\mu})~~~,
             \nonumber \\
 (-1)^{\mu} k_{t,\mu}   & =  & -
 \frac{1}{24\pi^2} \epsilon_{\mu\nu\rho\sigma} \int_{f(t,\mu )} d^{3}x
 tr[{\cal S}^{t}_{\nu}(x){\cal S}^{t}_{\rho}(x)
 {\cal S}^{t}_{\sigma}(x)] \nonumber \\
       &     + &
\frac{1}{8\pi^2} \epsilon_{\mu\nu\rho\sigma}
\int_{p(t+\nu,\mu,\nu)} d^{2}x
 tr[{\cal P}^{t}_{\rho}(x){\cal S}^{t}_{\sigma}(x)]~~~.
\end{eqnarray}
The actual expressions for $s^{n}_{t,\mu}(x)$ and
$p^{n}_{t+\nu,\mu\nu}(x)$ are given in ref.~\cite{Luesch}.
In Seibergs version we  use the same expressions but replace
\begin{equation}
 [s^{n}_{t,\mu}(x), p^{n}_{t+\nu,\mu\nu}(x), k_{t,\mu}]
\rightarrow
 [S_{t,\mu}(x), P_{t+\nu,\mu\nu}(x), K_{t,\mu}]~~~.
\end{equation}
The Seiberg charge will be gauge invariant, if we use the
same restriction as in two dimensions, and
the Chern-Simons term is given by $N_{CS} \equiv \sum_{n_s} K_{n_{s},\mu}$.
In the naive continuum limit $a \rightarrow 0$ one  finds
with $U(n,\mu)  = \exp(aA_{n,\mu})$
\begin{equation}
 (-1)^{\mu}K_{t,\mu} = - \frac{a^{3}}{8\pi^{2}}
 \epsilon_{\mu\nu\rho\sigma}
   tr[A_{t,\nu}(\partial_{\rho}A_{t,\sigma}+
   \frac{2}{3}A_{t,\rho}A_{t,\sigma})]~~.
\end{equation}
It is easy to show that
\begin{equation}
   K_{t,\mu} =  k_{t,\mu} +  K^{w}_{t,\mu}~~,
\end{equation}
where $K^{w}_{t,\mu}$ has the same form as $K_{t,\mu}$, but with
$S_{t,\mu}(x) \rightarrow   w^{t}(x)$.
As  the section transforms like a gauge
transformation this implies that the Chern-Simons number will change by an
integer.  For pure gauge fields one finds in addition
$s^{n}_{t,\mu}(x) = 1$ and
 $S_{t,\mu}(x) = P_{t+\nu,\mu\nu}(x)$ at the plaquette.
We, therefore, find
\begin{equation}
  (-1)^{\mu} K_{t,\mu} = -
 \frac{1}{24\pi ^2} \epsilon_{\mu\nu\rho\sigma}
          \int_{f(t,\mu )} d^{3}x
  tr[{\cal W}^{n}_{\nu}  {\cal W}^{n}_{\rho}  {\cal W}^{n}_{\sigma}],\;
   {\cal W}^{n}_{\nu} = w^{n}(x)^{-1}\partial_{\nu}w^{n}(x).
\end{equation}
This expression closely resembles the continuum form. After summation over the
spatial volume $N_{CS}$ indeed becomes an integer even on finite lattices.

We finally note that the topological charges, obtained from the
two approaches discussed above, are related \cite{Lau}.
Using eqs.(3.2) and (3.5) one has for any configuration
\begin{eqnarray}
q^{S}(n) & = & q^{L}(n) - q^{w}(n) \nonumber \\
q^{w}(n) & = &
 \frac{1}{24\pi ^2} \epsilon_{\mu\nu\rho\sigma}
 \int_{\partial c(n)} d^{3}x
  tr[{\cal W}^{n}_{\nu}  {\cal W}^{n}_{\rho}  {\cal W}^{n}_{\sigma}]~~~.
\end{eqnarray}
Here, $q^{w}(n)$ is the topological charge (integer) of the section and
$Q^{L} = Q^{w}$. Notice, that there is no restriction on $q^{L}(n)$
so that $q^{S}(n) = q^{L}(n)$ up to integers.
For smooth fields like instantons they always agree, while for realistic
field configurations this is true for almost every cell.

\section{Chern-Simons number in the strong coupling limit, finite size effects}

We will discuss here the distribution of CS-numbers induced by the Haar measure
on finite lattices,
{\it i.e.} in the absence of any Boltzmann weight factors in the
Euclidean path integral over configuration space. We will call this the strong
coupling limit. For the U(1) Higgs-model
in two dimensions it is easy to give an expression for the
distribution of Chern-Simons numbers in this limit.
Using the definition given in eq.(3.10) one finds on a lattice with spatial
extent $n$ the following recursive relation for the density,
$\rho_{n}(z)$, of CS-numbers $z$,
\begin{equation}
     \rho_{1}(z)   =   \left\{ \begin{array}{ll}
              1  & \mbox{$|z| \leq 1/2$} \\
              0  & \mbox{otherwise}
                           \end{array} \right.
\end{equation}
\begin{equation}
 \rho_{n}(z)  =  \prod_{i=1}^{n} \int_{-\frac{1}{2}}^{\frac{1}{2}}
                  dz_{i} \rho_{1}(z_{i})
               \delta (\sum_{j=1}^{n} z_{j} - z)
               =
                \int_{max(\frac{1-n}{2},z-\frac{1}{2})}
                    ^{min(\frac{n-1}{2},z+\frac{1}{2})}
                 dz \rho_{n-1}(z)~~~.
\end{equation}
Using this recursion formula we find explicitly,
\begin{eqnarray}
 \rho_{2n+1}(z) & = & \frac{1}{(2n)!} \sum_{i=0}^{n-m} (-1)^{i}
      \left( \begin{array}{c} 2n+1 \\ i \end{array} \right)
           (n+\frac{1}{2}-i-|z|)^{2n},\;
           m-\frac{1}{2}\leq |z| \leq m+\frac{1}{2} \nonumber \\
 \rho_{2n}(z) & = & \frac{1}{(2n-1)!} \sum_{i=0}^{n-1-m} (-1)^{i}
      \left( \begin{array}{c} 2n \\ i \end{array} \right)
           (n-i-|z|)^{2n-1},\;m\leq |z| \leq m+1.
\end{eqnarray}
These relations allow us to define moments of the distributions as
\begin{equation}
 {\cal P}_{n}^{\alpha}   =
 \int_{-\frac{n}{2}}^{\frac{n}{2}} dz z^{\alpha} \rho_{n} (z)~~.
\end{equation}
Some properties of $\rho_n(z)$ and these moments are discussed in the appendix.
In
particular it is obvious that all odd moments vanish
and  $\rho_n(z)/\rho_n(0) \rightarrow 1$ for $n \rightarrow \infty$.
However, on a lattice of finite extent $n$,
the distribution functions, $\rho_n$, are close
to Gaussians, {\it i.e.} large values of the CS-numbers are suppressed due to
finite size effects. This is reflected in the even moments,
\begin{eqnarray}
 {\cal P}_{n}^{0} & = & 1   ~~~,   \nonumber \\
 {\cal P}_{n}^{2} & = & \frac{n}{12}~~, \nonumber \\
 {\cal P}_{n}^{4} & = & -\frac{n}{120} + \frac{n^2}{48}~~.
\end{eqnarray}
The leading term for
 ~${\cal P}_{n}^{2\alpha} = (2\alpha - 1)!!({\cal P}_{n}^{2})^{\alpha}$
 ~is just coming~ from a ~Gaussian approximation\footnote{
The second moment has previously been evaluated in ref.~\cite{Wiese}.}.
Therefore $\langle N_{CS}^{2} \rangle$ grows with the number of links
in a timeslice.  Notice the fourth order cumulant
${\cal P}_{n}^{4}  -3 ({\cal P}_{n}^{2})^{2}  = -\frac{n}{120}$.

We will mainly be interested in the gauge invariant,
non-integer part of the CS-numbers. This can be obtained by summing over the
various gauge equivalent sectors of the distribution functions. On a
lattice with (2n+1) sites in spatial direction these are given by (see
Appendix eq.(A.1) for even number of lattice sites)
\begin{eqnarray}
R^{\alpha}_{2n+1}(z) & = & \sum_{m=-n}^{n} (z+m)^{\alpha} \rho_{2n+1}(z+m)
 ~~~,\;\; |z| \leq \frac{1}{2}~~~.
\end{eqnarray}
We show in the appendix that on a lattice of size $n$ the distribution
$R^0_n$ and the moments $R^{\alpha}_n$ are, in fact, independent of $z$
in the strong coupling limit, and equal the moments ${\cal
P}^{\alpha}_n$. In particular, this means that
$R^0_n(z) \equiv 1$ for all $z$, \ie~ the distribution of
the non-integer part of the CS-numbers is flat in configuration
space even on finite lattices.

Let us now discuss some numerical results obtained in the strong
coupling regime. We used a 2d-U(1) lattice Higgs model.
The lattice action is
\begin{eqnarray}
  S = & - & \frac{\beta}{r} \sum_{n,\mu <\nu}tr(U_{n,\mu\nu})
        +   \lambda \sum_{n}( \frac{1}{r}tr(
        \Phi^{\dagger}_{n}\Phi_{n})- 1)^2
      \nonumber \\
    & - & \frac{2\kappa}{r}
    \sum_{n,\mu}tr(\Phi^{\dagger}_{n}U_{n,\mu}\Phi_{n+\mu})
      +  \frac{1}{r} \sum_{n} tr(\Phi^{\dagger}_{n}\Phi_{n}),
\end{eqnarray}
with $r = 1$ for $U(1)$ and $r=2$ for $SU(2)$. We studied the distribution
of CS-numbers for various parameter sets on lattices of size $12^2$ and $24^2$.
In the left column of Fig.1
we show three distributions of the CS-numbers obtained from calculations
on the $12^2$ lattice at $(\beta ,\lambda ) = (2.0,\infty)$ and
$\kappa = (0.5,2.0,8.0)$. At each of the $\kappa$ values 100.000 measurements
of the CS-number, defined by eq.(3.10), in each of the 12 time-slices has
been performed.  The distributions for all 12 slices have then been added
in order to increase the statistics.

The $\beta$ value has been chosen such that our strong coupling calculations
should be applicable, whenever the Higgs sector decouples from the gauge field
sector. This happens at small values of $\kappa$. Indeed, no periodic structure
is visible in the distribution for our lowest value of $\kappa$ and the
distribution is well described by our analytic result, eq.(4.3).
At large values of
$\kappa$, however, the periodic structure becomes transparent. More and
more of the configurations are close to pure gauge, hence the clustering
of $N_{CS}$ around integers. We note, however, that the
peak heights are distributed according to
the characteristic Gaussian form of the strong coupling distribution
function. In the right column of Fig.1 we have divided the original
distributions by the strong coupling form of the distribution,
$\rho_{12}(z)$, which reflects the non-uniform distributions
of CS-numbers in phase space due to finite lattice effects.
The peak heights are now identical (The large fluctuations at higher
$N_{CS}$ values are due to limited statistics in this exponentially suppressed
part of the distribution.).
Although the shape of the distributions changes drastically, the moments
of $N_{CS}$ are insensitive to this as can be seen from Table 1.
For each $\kappa$ value we find on the $12^2$ lattice
$\langle N_{CS}^2 \rangle \simeq 1.00$ and
$\langle N_{CS}^4 \rangle \simeq 2.90$, which is
in perfect agreement with the exact results ${\cal P}_{12}^{2} = 1$
and ${\cal P}_{12}^{4} = 2.9$. Likewise
for the $24^2$ lattice, $\langle N_{CS}^2 \rangle  = 2.09$ and $\langle
N_{CS}^4 \rangle  = 11.80$
compared to ${\cal P}_{24}^{2} = 2$ and ${\cal P}_{24}^{4} = 11.8$.
We also have checked that these results are independent of $\beta$.
Hence, for large lattices the data are thus very well described by a Gaussian
distribution.

\begin{table}
\caption{Moments of the distribution of Chern-Simons numbers for the 2d-U(1)
Higgs model on a $12^2$ lattice at $(\beta,\lambda)=(2.0,\infty)$. The last
column shows the strong coupling results given in eq.(4.5).}
\vskip 20pt
\begin{center}
\begin{tabular}{|c|c|c|c|c|}
\hline
\rule{0pt}{13pt}$~~\kappa~~$
                               &    0.5    &    2.0   &    8.0    & 0.0 \\
\hline
\rule{0pt}{13pt}$\langle N_{CS}^2 \rangle$
                               & 0.998(5)  & 1.001(8) & 0.970(13) & 1.0 \\
\hline
\rule{0pt}{13pt}$\langle N_{CS}^4 \rangle$
                               & 2.886(21) & 2.909(34) & 2.830(90) & 2.9 \\
\hline
\end{tabular}
\end{center}
\end{table}

The results presented so far are in accordance with the assumption that
the distribution of CS-numbers on a lattice of size $n_{\sigma} \times
n_{\tau}$ is a product of two probability
distributions, where one is just given by $\rho_{n_\sigma}$ and
describes
the phase space restrictions for $N_{CS}$ and the other gives the
probability to find a certain non-integer part, $z$, for the
CS-number, $N_{CS} = m+z$. This part of the distribution is
controlled by the action and thus contains the relevant physical
information. Let us write the probability to find CS-number
$N_{CS} = m+z$ as
\begin{eqnarray}
P_{n_{\sigma},n_{\tau}} (N_{CS}) = \omega_{n_{\sigma},n_{\tau}} (z)
\rho_{n_{\sigma}} (m+z)~~~|z|\le{1 \over 2}~~~,
\end{eqnarray}
with $\rho_{n_{\sigma}}$ defined in eq.(4.3) and
$\omega_{n_{\sigma},n_{\tau}}$ denoting a
probability distribution that depends on the couplings of the Euclidean
action as well as the spatial and temporal size of the lattice. Using
eq.(4.6) and (A.13) it is
then easy to verify that indeed the moments, $\langle N_{CS}^{\alpha}
\rangle$, are independent of $\beta$ and $\kappa$.

Let us now discuss the distribution of the non-integer part of the
CS-number and the corresponding moments. These describe the fluctuations
of the CS-number around vacuum configurations
in a given topological sector. In
the continuum theory arguments have been given that
these vacuum fluctuations are also proportional to
the spatial volume of the system \cite{Shap}. In Table 2 we give results
for the moments of the non-integer part of the CS-numbers. We note that
for a flat distribution one finds $\langle z^2\rangle = 1/12$ and
$\langle z^4 \rangle = 1/80$. These limiting values agree with our
results at $\kappa =0.5$ and are also approached for larger values of
$\kappa$ with increasing size of the lattice.

\begin{table}
\caption{Moments of the distributions of the
non-integer part ($z$) of Chern-Simons numbers, $N_{CS} = m +z$,
for the 2d-U(1)
Higgs model on a $12^2$ and $24^2$ lattices at $(\beta,\lambda)=(2.0,\infty)$.
}
\vskip 20pt
\begin{center}
\begin{tabular}{|c|c|c|c|c|}
\hline
\rule[1pt]{0pt}{13pt}$~~\kappa~~$ &
$\langle z^2 \rangle~,~n=12$ & $\langle z^2 \rangle~,~n=24$ &
$\langle z^4 \rangle~,~n=12$ & $\langle z^4 \rangle~,~n=24$ \\
\hline
\rule[1pt]{0pt}{13pt}0.5 & 0.0832(1) & 0.0833(1) & 0.0125(1) & 0.0125(1) \\
                     2.0 & 0.0662(1) & 0.0805(1) & 0.0091(1) & 0.0119(1) \\
                     8.0 & 0.0287(1) & 0.0511(1) & 0.0024(1) & 0.0062(1) \\
\hline
\end{tabular}
\end{center}
\end{table}

\section{Monte-Carlo results in 4d}

Before discussing the results for the 4d-SU(2) Higgs model, it is necessary to
give some details of the programs developed by us to evaluate  $K_{t,\mu}$.
Because the interpolation in the Seiberg case is done on the original
links and not on the gauge fixed links, the fields tend to be very
rough. The same happens if one wants to evaluate the topological
charge $q^{w}(n)$ via the section.
One of the integrations in eq.(3.12) can be done in analytic form,
so we are left with a two dimensional integral. These integrals must
be evaluated carefully, if one wants to extract the periodic structure in the
distribution of CS-numbers.
On the other hand, if we are only interested in the non-integer and
therefore gauge invariant
part of $N_{CS}$ (used in the discussion of baryon number violation)
then it suffices to perform several Landau
gauge fixing sweeps before doing the integrals. The effect is that
$N_{CS}$ will be shifted by an integer, in most cases into the interval
$-1/2 \le N_{CS} \le 1/2$. The advantage is that gauge fields on
the links become very smooth and the integrals will converge much
faster. To define a convergence criterion we note that $K_{t,\mu}$
must change by an integer under a gauge transformation. We took
various configurations and monitored $N_{CS}$, first without and then
with a Landau gauge fixing sweep.
If $\delta K_{n,\mu}$ was an integer within a relative
error of $\epsilon = 10^{-4}$, the integrals were accepted.

We have used the following strategy, which turned out to be very
efficient. The volume term in eq.(3.12) is the most time consuming part, so we
concentrate on that. Since the Chern-Simons numbers on each timeslice
are hardly correlated we can use all of them.
We perform a Gauss-Legendre integration with $8\times 8$ points and
store the results for all the  $K_{n,4}$'s in each timeslice. We then repeat
this calculation with $16\times 16$ points and compare the results
for each $K_{n,4}$. If the relative
difference is less than $\epsilon$ the contribution is accepted.
Otherwise we collect the $K_{n,4}$'s which have not yet converged.
In these cells we repeat the integration with $32\times 32$ points instead,
compare with the previous values and eventually repeat the procedure with
$64\times 64$ points. This part can be done very efficiently in vectorized
form.  At this point only in a few cells the results for $K_{n,4}$'s have
not converged, for these we use a library integration
routine with interval adaptation. The typical time for evaluation of
$N_{CS}$ (without any gauge fixing)
in a time slice with $6^3$ lattice points is roughly
60 seconds on the CRAY-YMP. With gauge fixing this is reduced to
about 6 seconds. The entire program runs with a speed of about
(150-200)Mflops.

As a test of our programs we calculated the time dependence of $N_{CS}$ on a
4d - SU(2) instanton configuration in the complete axial gauge.
This is shown in Fig.2 for an instanton
of core size $\rho = 2$, on an $8^3 \times 12$ lattice. On the first
timeslice the gauge fields are in the vacuum sector with
$N_{CS} = 0$. On the last timeslice the instanton has mediated the
tunneling to
the other vacuum sector with $N_{CS} = 1$. We note that in the middle
($t=6$) one finds $N_{CS} = 1/2$. The corresponding field
configuration on this timeslice is the analog to the sphaleron in the Higgs
models and interpolates between the two vacuums at $t=1$ and $t=12$.

For the  SU(2) gauge Higgs model we used the action given in
eq.(4.7) with $r=2$. We have chosen to study the structure of the
CS-number distributions for various values of $\kappa$ at fixed
$(\beta,\lambda) = (2.25,0.5)$. For this choice of parameters the
phase transition between the symmetric and broken phases is known to
occur at $\kappa \simeq 0.27$ \cite{Bock}.
We have performed simulations on a $4^3\times 2$ lattice at
$\kappa = 0.25,$ 0.30 and 0.40, {\it i.e.} on both sides of the transition
line. For each of the $\kappa$ values we have measured $N_{CS}$ on
2000 gauge field configurations without any
gauge fixing. In  the left column of Fig.3
we show the CS-number distributions. The periodic
structure is obvious even for $\kappa$ below $\kappa_{c}$.
With increasing $\kappa$ the CS-numbers on many of the configurations
are close to an integer, and therefore the gauge field configurations
can be interpreted as being close to pure gauge.
We checked that $N_{CS}$ indeed changed by an integer under a gauge
transformation, which confirms the quality of our numerical
integrations. At $\kappa =0.3$ we have performed axial as well as Landau
gauge fixing. Even after a few gauge fixing sweeps nearly all CS-numbers
get shifted into the interval $[-1/2,1/2]$. The distribution after only
a single gauge fixing sweep is shown in
Fig.4. With a few more Landau gauge fixing sweeps
only the peak in the middle would remain.

For the moments of $N_{CS}$ we find on
the $4^3 \times 2$ lattice at $\kappa = 0.30$ the values
$\langle N_{CS}^2 \rangle = 2.93$  and $\langle N_{CS}^4 \rangle = 24.57$.
We note that a Gaussian fit with width 2.93 would suggest
$\langle N_{CS}^4 \rangle = 24.96$.
Like in two dimensions the moments are thus very close to those of Gaussian
distributions. Moreover, we also find for the 4d-SU(2) Higgs model that
the moments are independent of $(\beta ,\kappa)$ and increase
proportional to the spatial volume. We have performed simulations on a
$6^3\times 2$ lattice (without gauge fixing) at
$(\beta,\lambda) = (2.25,0.5)$ and $\kappa = (0.30,0.40)$.
For this lattice we find $\langle N_{CS}^2 \rangle = 9.63$ and
$\langle N_{CS}^4 \rangle = 274.39$. A Gaussian approximation would lead
to  $\langle N_{CS}^4 \rangle = 275.26$. The general
structure of the distributions thus is very similar to the 2d-U(1) case
and we expect that the
finite size effects can be eliminated similarly by dividing out the
distribution induced solely by the Haar measure (strong coupling
distribution). We approximate this by a Gaussian distribution with the
width given by our numerically determined value for $\langle N_{CS}^2
\rangle$.
This is shown in the right column of Fig.3.
The peaks are now more or less of equal height. We thus may expect, that
similar to the 2d case the finite size effects in the distributions drop
out in the distribution functions $\omega_{n_{\sigma},n_{\tau}} (z)$ for
the gauge invariant, non-integer
part, $z$, defined in eq.(4.8). We note that these can be calculated
on gauge fixed configurations, which is computationally much less
demanding.

\noindent
At finite temperature the tunneling between topologically distinct vacuums
should become more likely with increasing temperature. This should be
reflected in a flattening of the distributions,
$\omega_{n_{\sigma},n_{\tau}} (z)$, defined in eq.(4.8). In order
to see, whether this effect is visible in our distributions, we have
performed simulations on lattices of size
$6^{3}\times n_{\tau}$ with $n_{\tau} = 2$, 4 and 6.
For these lattices we used 50 Landau gauge fixing
sweeps, to make the integrals converge fast. For both lattice sizes
we collected 6000 Chern-Simons numbers at $\kappa=0.3$.
The results for $\omega_{n_{\sigma},n_{\tau}} (z)$ are shown in
Fig.5. The left column shows the change in the distributions with
varying $n_{\tau}$ at fixed $n_{\sigma}$. The tendency for a
flattening of the distribution at finite temperature is clearly visible.
We take this as evidence that the system tunnels more often at non-zero
temperature.

A major problem for a more quantitative analysis of the distributions
at finite temperature is caused by the volume dependence of
$\omega_{n_{\sigma},n_{\tau}} (z)$ itself, which leads to a flattening
of the distributions with increasing spatial volume. This is seen in
the right column of Fig.5, where we compare
distributions on $4^3 \times 2$, $6^3 \times 2$
and $8^3 \times 2$ lattices. Clearly the distributions become flatter
with increasing spatial lattice size. We note, that this effect,
caused by the spatial volume, is opposite to the finite temperature
effect shown in the left column of Fig.5. In that case
the distributions become more peaked around $N_{CS}=0$,
although the total 4-volume increases.

\begin{table}
\caption{The width of the distribution of the
non-integer part ($z$) of Chern-Simons numbers, $N_{CS} = m +z$,
for the 4d-SU(2)
Higgs model on lattices of size $n_{\sigma}^3 \times n_{\tau}$
at $(\beta,\lambda)=(2.25,0.5)$ and various values of $\kappa$.
}
\vskip 20pt
\begin{center}
\begin{tabular}{|c|c|c|c|}
\hline
\rule[1pt]{0pt}{13pt}$~~N_{\sigma}~~$ & $~~N_{\tau}~~$ &
$~~\kappa~~$ & $~~\langle z^2 \rangle~~$ \\
\hline
\rule{0pt}{13pt}4 & 2 & 0.30  & 0.028(2)  \\
                6 & 2 & 0.30  & 0.068(2)  \\
                8 & 2 & 0.30  & 0.083(1)  \\
\hline
\rule{0pt}{13pt}6 & 2 & 0.30  & 0.068(2)  \\
                6 & 4 & 0.30  & 0.049(2)  \\
                6 & 6 & 0.30  & 0.049(2)  \\
\hline
\rule{0pt}{13pt}4 & 2 & 0.25  & 0.057(2)  \\
                4 & 2 & 0.30  & 0.028(2)  \\
                4 & 2 & 0.40  & 0.009(2)  \\
\hline
\end{tabular}
\end{center}
\end{table}

The moments, $\langle z^2 \rangle$, of the distributions shown in
Fig.5 are summarized in Table 3. They show that the limiting value
for a flat distribution, $\langle z^2 \rangle = 1/12$ is rapidly
approached with increasing spatial volume as wells as decreasing
couplings $\beta$ and/or $\kappa$.  As mentioned before
the broadening of the distribution with increasing spatial
volume is not unexpected. It has been shown \cite{Shap} that even at zero
temperature quantum fluctuations of the fields can lead to large values
of $N_{CS}$. In fact, this contribution is proportional to the spatial
volume. In the present parameter range studied by us such a linear
dependence is not visible. We will have to work at much large values
of the couplings $\beta$ and/or $\kappa$.
This will be necessary in order to proceed with a
quantitative analysis of finite temperature
effects on the CS-number distributions.

\section{Conclusions}
We have studied the properties of Chern-Simon numbers in two and four
dimensional Higgs models on Euclidean lattices. We have shown that
the definition of the Chern-Simons term based on the geometric
definition of topological charges preserves the basic properties of the
continuum expressions. The non-integer part of the Chern-Simons number is
gauge invariant; the Chern-Simons number changes by an integer under
gauge transformations. Moreover, we have shown that the effective
potential for the Chern-Simons number is periodic with maxima of equal
height at integer values of $N_{CS}$, if finite size effects are taken
into account properly.

The distribution of Chern-Simon numbers flattens with increasing
temperature. This will lead to an increase of baryon number violating
processes at high temperature. At present we could, however, not explore
the temperature of the transition rates quantitatively as the
distributions are still influenced by finite size effects caused by
quantum fluctuations in the trivial vacuum. These fluctuations will
be suppressed at larger values of the gauge coupling, which we might
have to choose for a future quantitative analysis of the Chern-Simons
number distributions in the vicinity of the symmetry restoring phase
transition of the 4d-SU(2) Higgs model \cite{Bunk}. Another
possibility to suppress the contribution from
vacuum fluctuation is based on a modification of the cooling method,
which also allows to smoothen configurations and obtain the contribution of
classical configurations that extremize the Euclidean action \cite{Duncan}.
This approach has recently been used in the context
of the 2d O(3) nonlinear sigma model \cite{Kripfganz} and is in an gauge
invariant manner also applicable to SU(2) gauge theories \cite{Sijs} .
It will be interesting to test with our geometric formulation
whether the Chern-Simons numbers develop a plateau
under extremization of the Euclidean action. This would allow to extract
the distribution of Chern-Simons numbers on topologically
non-trivial gauge field configurations for which the quantum
fluctuations are suppressed.

\section*{Acknowledgement}
The numerical simulations described here have been performed on the
Cray-YMP at HLRZ. Financial support from DFG under contract Pe 340/1-3
and the Ministerium f\"ur Wissenschaft und Forschung NRW under contract
IVA5-10600990 is gratefully acknowledged.
\section*{Appendix}
\renewcommand{\theequation}{A.\arabic{equation}}

We will first show that the following identities are true on lattices
with spatial extent $(2n+1)$ and $2n$, respectively,
\begin{eqnarray}
\sum_{m=-n}^{n} \rho_{2n+1}(z+m) & = &1
 ~~~,\;\; |z| \leq \frac{1}{2}.
\nonumber \\
\sum_{m=-n}^{n-1} \rho_{2n}(z+m) & = &1
,\;\; 0 < z \leq 1~~~.
\end{eqnarray}
One has
\begin{eqnarray}
 (2n)! \sum_{m=-n}^{n} \rho_{2n+1}(z+m) & = &
  \sum_{m=1}^{n} \sum_{i=0}^{n-m} (-1)^{i}
      \left( \begin{array}{c} 2n+1 \\ i \end{array} \right)
           (n+\frac{1}{2}-i-z-m)^{2n} \nonumber \\
           & + &
  \sum_{m=1}^{n} \sum_{i=0}^{n-m} (-1)^{i}
      \left( \begin{array}{c} 2n+1 \\ i \end{array} \right)
           (n+\frac{1}{2}-i+z-m)^{2n}  \nonumber \\
           & + &
  \sum_{i=0}^{n} (-1)^{i}
      \left( \begin{array}{c} 2n+1 \\ i \end{array} \right)
           (n+\frac{1}{2}-i-|z|)^{2n}.
\end{eqnarray}
Interchanging the summation over $(i,m)$ in the double sum gives
for the first term
\begin{eqnarray}
&&\sum_{m=1}^{n} \sum_{i=0}^{n-m} (-1)^{i}
      \left( \begin{array}{c} 2n+1 \\ i \end{array} \right)
           (n+\frac{1}{2}-i-z-m)^{2n}  \nonumber \\
           & = &
  \sum_{i=0}^{n-1} \sum_{m=0}^{i} (-1)^{m}
      \left( \begin{array}{c} 2n+1 \\ m \end{array} \right)
           (n-\frac{1}{2}-i-z)^{2n}  \nonumber
 ~~~.
\end{eqnarray}
The summation over $m$ can now be performed giving for the first term in
eq.(A.2)
\begin{eqnarray}
  \sum_{i=0}^{n-1} (-1)^{i}
      \left( \begin{array}{c} 2n \\ i \end{array} \right)
           (n-\frac{1}{2}-i-z)^{2n}~~~.
\end{eqnarray}
Using now also the recursion relation,
\begin{equation}
      \left( \begin{array}{c} 2n+1 \\ i \end{array} \right)  =
      \left( \begin{array}{c} 2n \\ i \end{array} \right) +
      \left( \begin{array}{c} 2n \\ i-1 \end{array} \right)~~~,
\end{equation}
in the third term of eq.(A.2), we obtain
\begin{equation}
 (2n)! \sum_{m=-n}^{n} \rho_{2n+1}(z+m)  =
  \sum_{i=0}^{2n} (-1)^{i}
      \left( \begin{array}{c} 2n \\ i \end{array} \right)
           (n-\frac{1}{2}-i+z)^{2n} = (2n)!~~~~.
\end{equation}
This concludes the proof.

Let us now calculate the moments of this distribution.
We consider the integrals
\begin{eqnarray}
 {\cal P}_{2n}^{\alpha} & = & \int_{-n}^{n} dz z^{\alpha} \rho_{2n}
                          \nonumber \\
              & = & \frac{2}{(2n-1)!} \sum_{m=0}^{n-1}
                                      \sum_{i=0}^{n-1-m} (-1)^{i}
      \left( \begin{array}{c} 2n \\ i \end{array} \right)
                     \int_{m}^{m+1} dz z^{\alpha} (n-i-|z|)^{2n-1}~.
\end{eqnarray}
Interchanging the summation over $(i,m)$ in the double sum
and then integrating over z yields
\begin{equation}
 {\cal P}_{2n}^{\alpha}
                =   \frac{\alpha !}{(2n+\alpha)!}
                                      \sum_{i=0}^{2n} (-1)^{i}
      \left( \begin{array}{c} 2n \\ i \end{array} \right)
                     (n-i)^{2n+\alpha}~~~.
\end{equation}
The Stirling numbers of the second kind,
\begin{equation}
 {\cal S}_{n}^{m}  =   \frac{(-1)^m}{m!}\sum_{i=0}^{m} (-1)^{i}
      \left( \begin{array}{c} m \\ i \end{array} \right) i^{n}~~~,
\end{equation}
with the recursion
\begin{equation}
 {\cal S}_{n}^{m}  = \sum_{i=1}^{m}i {\cal S}_{i+n-m-1}^{i},\;\;
 {\cal S}_{m}^{m} = 1
\end{equation}
can be used to write the moments as
\begin{equation}
 {\cal P}_{2n}^{\alpha} =   \frac{(2n)!\alpha !}{(2n+\alpha)!}
          \sum_{i=0}^{\alpha} (-n)^{\alpha - i}
      \left( \begin{array}{c} 2n+\alpha \\ \alpha - i \end{array} \right)
                                     {\cal S}_{2n+i}^{2n}~~~.
\end{equation}
In the following we shall only consider the cases $\alpha = (0,1,2,3,4)$.
{}From the recursion formula it follows easily
\begin{eqnarray}
 {\cal S}_{m+1}^{m} & = &
  \left( \begin{array}{c} m+1 \\ 2 \end{array} \right) ~~~,\nonumber \\
 {\cal S}_{m+2}^{m} & = & \frac{3m+1}{4}
  \left( \begin{array}{c} m+2 \\ 3 \end{array} \right) ~~~,\nonumber \\
 {\cal S}_{m+3}^{m} & = & \frac{m(m+1)}{2}
  \left( \begin{array}{c} m+3 \\ 4 \end{array} \right) ~~~,\nonumber \\
 {\cal S}_{m+4}^{m} & = & \frac{15m^{3}+30m^{2}+5m-2}{48}
  \left( \begin{array}{c} m+4 \\ 5 \end{array} \right)~~~.
\end{eqnarray}
This leads to
\begin{eqnarray}
 {\cal P}_{2n}^{0} & = & 1   ~~~, \nonumber \\
 {\cal P}_{2n}^{1} & = & 0   ~~~, \nonumber \\
 {\cal P}_{2n}^{2} & = & \frac{n}{6}~~~, \nonumber \\
 {\cal P}_{2n}^{3} & = & 0   ~~~,   \nonumber \\
 {\cal P}_{2n}^{4} & = & -\frac{n}{60} + \frac{n^2}{12}~~~.
\end{eqnarray}

We also have checked that the following relations for the
"discrete" momenta hold as long as the degree of the moment does not
exceed the lattice size,
\begin{eqnarray}
\sum_{m=-n}^{n} (z+m)^k \rho_{2n+1}(z+m) & = &{\cal P}^k_{2n+1}~,
 ~k < 2n+1~,\;\; |z| \leq \frac{1}{2}.
\nonumber \\
\sum_{m=-n}^{n-1} (z+m)^k \rho_{2n}(z+m) & = &{\cal P}^k_{2n}~,
 ~k<2n~,\;\; 0 < z \leq 1~~~.
\end{eqnarray}

\end{document}